%% file: sigir.tex
\documentclass[sigconf]{acmart}
\AtBeginDocument{%
  }

\copyrightyear{2026}
\acmYear{2026}
\setcopyright{cc}
\setcctype{by}
\acmConference[SIGIR '26]{Proceedings of the 49th International ACM SIGIR Conference on Research and Development in Information Retrieval}{July 20--24, 2026}{Melbourne, VIC, Australia}
\acmBooktitle{Proceedings of the 49th International ACM SIGIR Conference on Research and Development in Information Retrieval (SIGIR '26), July 20--24, 2026, Melbourne, VIC, Australia}
\acmDOI{10.1145/3805712.3808420}
\acmISBN{979-8-4007-2599-9/2026/07}
\begin{document}


\title[Unified Supervision for Sponsored Retrieval at Walmart]{Unified Supervision for Walmart's Sponsored Search Retrieval via Joint Semantic Relevance and Behavioral Engagement Modeling}

\author{Shasvat Desai}
\orcid{0000-0002-4001-6467}
\authornote{These authors contributed equally to this work.}
\email{shasvat.desai@walmart.com}
\affiliation{
  \institution{Walmart Global Tech}
  \city{Hoboken}
  \country{USA}
}

\author{Md Omar Faruk Rokon}
\orcid{0000-0002-1385-9389}
\authornotemark[1]
\email{mdomarfaruk.rokon@walmart.com}
\affiliation{
  \institution{Walmart Global Tech}
  \city{Sunnyvale}
  \country{USA}
}

\author{Jhalak Nilesh Acharya}
\orcid{0009-0005-6686-6055}
\email{jhalak.acharya@walmart.com}
\affiliation{
  \institution{Walmart Global Tech}
  \city{Sunnyvale}
  \country{USA}
}

\author{Isha Shah}
\orcid{0009-0001-6498-7954}
\email{Isha.Shah@walmart.com}
\affiliation{
  \institution{Walmart Global Tech}
  \city{Sunnyvale}
  \country{USA}
}

\author{Hong Yao}
\orcid{0009-0007-9988-0069}
\email{hong.yao0@walmart.com}
\affiliation{
  \institution{Walmart Global Tech}
  \city{Sunnyvale}
  \country{USA}
}

\author{Utkarsh Porwal}
\orcid{0000-0003-2768-0184}
\email{utkarsh.porwal@gmail.com}
\affiliation{
  \institution{Walmart Global Tech}
  \city{Sunnyvale}
  \country{USA}
}

\author{Kuang-chih Lee}
\orcid{0009-0007-5198-9866}
\email{Kuangchih.Lee@walmart.com}
\affiliation{
  \institution{Walmart Global Tech}
  \city{Sunnyvale}
  \country{USA}
}

\renewcommand{\shortauthors}{Shasvat Desai et al.}

\begin{abstract}

Modern search systems rely on a fast first-stage retriever to fetch relevant items from a massive catalog of items. Deployed search systems often use user engagement signals (clicks, etc.) to supervise bi-encoder retriever training at scale, because these signals are continuously logged from real traffic and require no additional annotation effort. However, engagement is an imperfect proxy for semantic relevance: items may receive interactions due to popularity, promotion, attractive visuals, titles, or price, despite weak query-item relevance. These limitations are further accentuated in Walmart’s e-commerce sponsored search. User engagement on ad items is often structurally sparse because the frequency with which an ad is shown depends on factors beyond relevance—whether the advertiser is currently running that ad, the outcome of the auction for available ad slots, bid competitiveness, and advertiser budget. Thus, even highly relevant query–ad pairs can have limited engagement signals simply due to limited impressions. Moreover, e-commerce search pages typically allocate fewer slots for ads than for non-sponsored results, further limiting impressions and reducing engagement coverage over the candidate ad item set.

We propose a bi-encoder training framework for Walmart sponsored search retrieval in e-commerce that uses semantic relevance as the primary supervision signal, with engagement used only as a preference signal among relevant items. Concretely, we construct a context-rich training target by combining (i) graded relevance labels from a cascade of cross-encoder teacher models, (ii) a multi-channel retrieval prior score derived from the rank positions and cross-channel agreement of retrieval systems running in production, and (iii) user engagement applied only to semantically relevant items to refine preferences. Our approach outperforms the current production system in both offline evaluation and online A/B tests, yielding consistent gains in average relevance and NDCG.

\end{abstract}

\ccsdesc[500]{Applied computing~E-commerce infrastructure}
\ccsdesc[500]{Information systems~Novelty in information retrieval}
\keywords{Semantic Relevance in Bi-Encoder Retrieval; Large Scale E-commerce Search; User Engagement Aware Retrieval; Hard Negative Mining}

\maketitle

\input{Sections/Introduction}

\input{Sections/Related_Work}

\input{Sections/Multi_signal_data_curation}

\input{Sections/Evaluation}

\input{Sections/Discussion}

\bibliographystyle{ACM-Reference-Format}
\balance
\bibliography{sigir}
\clearpage
\section*{Author Biography}
\textbf{Shasvat Desai} is a Staff Data Scientist at Walmart Global Tech, specializing in applied machine learning for information retrieval and sponsored search. His work spans query and item understanding, intent-aware representations for retrieval and relevance, and ad user experience initiatives such as product highlights, sponsored questions, and themed carousels. Previously, he worked on computer vision systems for retail, including multi-camera scene understanding, object detection and tracking, and real-time edge deployment. He focuses on building scalable retrieval and ranking systems that balance relevance, engagement, diversity, and revenue under real-world latency and business constraints.

\vspace{0.5em}
\noindent
\textbf{Md Omar Faruk Rokon} is a Staff Data Scientist at Walmart specializing in applied machine learning for Information Retrieval. He earned his Ph.D. in Computer Science in 2022, with a research focus on embedding and representation learning. His current work leverages Transformer architectures and Large Language Models to enhance semantic understanding and search relevance. Additionally, he develops rigorous search evaluation methodologies to drive improvements in e-commerce systems

\vspace{0.5em}
\noindent
\textbf{Jhalak Nilesh Acharya} is a Data Scientist at Walmart's Ads Data Science team, specializing in relevance, ranking, and ad experiences. She completed her MS in Data Science in 2022, building upon her foundation of BE in Electronics and Telecommunication. Her current work focuses on developing innovative carousel experiences and leveraging machine learning, computational linguistics and data science techniques to optimize ad business metrics and enhance user engagement in e-commerce advertising systems.

\vspace{0.5em}
\noindent
\textbf{Isha Shah} is a Data Scientist at Walmart Global Tech based in San Francisco, California, working in AdTech Data Science. She specializes in computational advertising and large-scale information retrieval and ranking. Her current work focuses on optimizing sponsored ads systems, specifically developing advanced bidding strategies and unified supervision frameworks for ad retrieval and ads ranking. She holds an M.S. in Computer Science from Georgia Tech. Her research interests include machine learning for e-commerce, auction mechanisms, retrieval and ranking efficiency.

\vspace{0.5em}
\noindent
\textbf{Hong Yao} is a Distinguished Data Scientist at Walmart Global Tech, where his research focuses on improving advertising performance by delivering the right ads to the right users at the right time. He develops advanced ad targeting, retrieval, and ranking models to enhance personalization and effectiveness.
He has been a Senior Member of IEEE since 2009. Dr. Yao earned his Ph.D. in Computer Science from the University of Regina, specializing in machine learning and uncertainty reasoning.

\vspace{0.5em}
\noindent
\textbf{Utkarsh Porwal} is a Director at Walmart Global Tech. He leads agentic AI initiatives for B2B use cases. His work focuses on applying machine learning to large-scale information retrieval and advertising systems, with an emphasis on relevance, user experience, and monetization under real-world latency and business constraints.

\vspace{0.5em}
\noindent
\textbf{Kuang-chih Lee} is a Senior Director at Walmart Global Tech, where he leads research and development for real-time personalized e-commerce marketplaces. His research interests span search and recommendation systems, online advertising, fraud detection, supply chain optimization, inventory forecasting, dynamic pricing, and large-scale machine learning systems, including NLP and computer vision. He has published over 30 papers in top conferences and journals, including CVPR, NeurIPS, AAAI, CIKM, KDD, IEEE TPAMI, and CVIU, and holds more than 20 patents. His work has received over 5,000 citations according to Google Scholar. He received his Ph.D. in Computer Science from the University of Illinois Urbana–Champaign in 2005.
\end{document}

%% file: Sections/Introduction.tex
\section{Introduction}
\label{sec:intro}

Several deployed e-commerce search systems leverage user engagement signals (e.g., clicks, orders) to supervise retrieval training \cite{shi2023everyone, he2023que2engage, li2021embedding, fan2025synergizing}. However, engagement does not always reflect relevance: beyond query–item match, user interactions are shaped by item popularity, promotion, appealing images, flashy titles, and price. Prior work highlights two failure modes of engagement-based supervision: (i) engagement-derived labels introduce systematic noise, biasing retrievers toward highly engaged but weakly relevant items; and (ii) engagement is sparse for much of the candidate set, especially for cold-start and long-tail items, limiting learnability from engagement alone \cite{fan2019mobius, lin2024enhancing, zhang2022uni}.

To mitigate these issues, \cite{zhang2022uni, lin2024enhancing, fan2019mobius} incorporate explicit relevance signals, but typically in a secondary role—e.g., filtering engagement-derived positives with a relevance threshold—while engagement remains the primary supervision source that defines training pairs and drives optimization. In other words, relevance mainly prunes obvious false positives, but learning is still engagement-driven.

\begin{figure*}
    \includegraphics[width=0.95\textwidth]
    {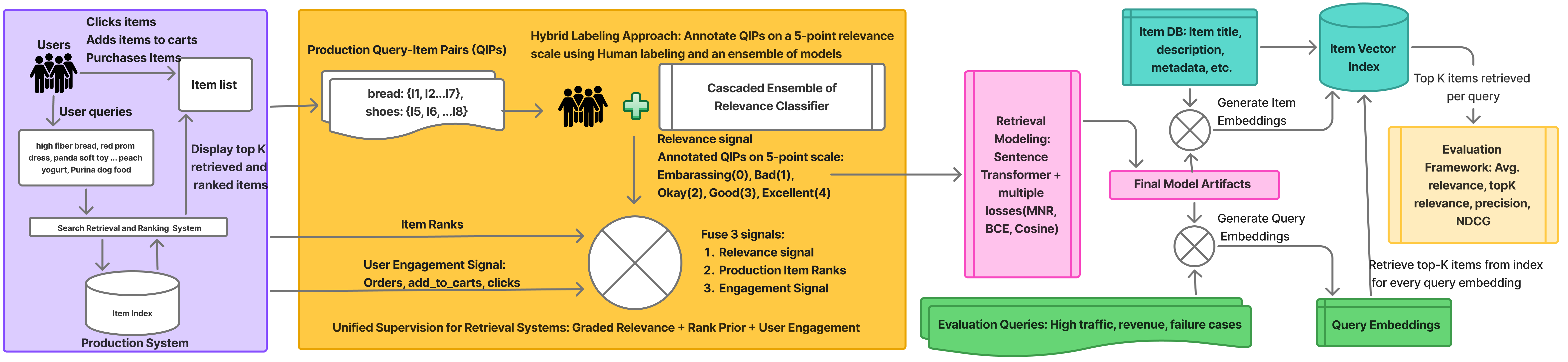}
    \caption{Overview of the unified supervision framework for large-scale e-commerce retrieval. The system integrates human and relevance cross encoder labeled relevance scores, item ranks from existing retrieval channels, and user engagement signals to generate a context-rich supervision target to train bi-encoder model.}
\end{figure*}

In our sponsored search setting at Walmart, this strategy is often insufficient for two key reasons. (i) Engagement is not only noisy but also structurally sparse because whether an ad receives impressions depends on factors beyond relevance—whether the advertiser is currently running the ad, the auction for available ad slots, bid competitiveness, and advertiser budget. Consequently, many relevant query–ad pairs receive too few impressions to yield stable engagement supervision, and observed engagement can reflect how the system selects and positions ads as much as semantic match. (ii) On Walmart search pages, fewer slots are available for ads than for non-sponsored results, further reducing engagement coverage over the candidate ad set.
Thus, we propose a unified supervision framework for sponsored retrieval at Walmart that makes semantic relevance the primary training signal and incorporates engagement only as a preference signal among relevant items. This design yields a context-rich target for training bi-encoder retrievers that is robust to sparse and exposure-biased engagement. \textbf{We contribute:}

\begin{itemize}
    \item We provide relevance-primary supervision using graded semantic relevance from cross-encoder teacher models.
    \item We make the retriever engagement-aware, aligning it with downstream ranking objectives.
    \item We combine results across multiple production retrieval channels to identify hard negatives (top-ranked but irrelevant items) and reduce engagement-driven false positives.
    \item We propose a unified supervision framework that combines these three signals into a single training target for bi-encoder retrieval in Walmart's sponsored search framework.

\end{itemize}

%% file: Sections/Related_Work.tex
\section{Related Work}
\label{sec:related}

Bi-encoder retrievers enable low-latency retrieval by independently encoding queries and items in embedding space \cite{karpukhin2020dpr}. A growing body of deployed search systems leverages user engagement signals (e.g., clicks and purchases) to supervise retrieval training at scale \cite{shi2023everyone, fan2025synergizing, he2023que2engage, pang2025generative, jha2024unified, li2021embedding}. However, engagement is an imperfect indicator of relevance, as shown by \cite{zhang2022uni, lin2024enhancing, fan2019mobius}. These works introduce relevance signals, typically as a secondary constraint— e.g, filtering engagement-derived positives with a relevance threshold—to reduce engagement-driven false positives. In practice, they still construct positives primarily from engaged query–item pairs. In Walmart sponsored search, however, engagement is structurally sparse because impressions are limited by ad-slot auctions, advertiser bids, and budgets. We therefore use semantic relevance to define positives and incorporate engagement only afterward as a preference signal among relevant items. Beyond deployed search systems, many research settings train retrieval models without large-scale engagement logs and instead rely on relevance supervision. A standard approach is to distill ranking knowledge from high-capacity cross-encoder teacher models into a bi-encoder \cite{izacard2021reader2retriever, lin2020distilling, hofstatter2020improving, lu2022ernie}, typically combined with hard-negative mining(HNM) to surface challenging irrelevant examples. \cite{ren2021rocketqav2, yu2023prod}.

Overall, prior work has improved bi-encoder relevance using distillation, HNM, or engagement supervision, often as separate mechanisms or with a single dominant signal defining training targets. In contrast, we propose a unified supervision framework by combining: (i) graded relevance labels from cross encoder teacher models, (ii) a multi-channel retrieval prior score derived from production retrieval channels, and (iii) engagement. This unified target makes semantic relevance the foundation of training while injecting engagement and production-channel evidence, and to the best of our knowledge is the first sponsored-search retrieval framework to integrate all three signals into a single bi-encoder training objective.

%% file: Sections/Multi_signal_data_curation.tex

\section{Unified Supervision in Bi-Encoder 
}
\label{sec:msd}

We denote a query--item pair as $(q,i)$, and refer to it as a QIP. We propose a unified supervision framework that synthesizes target score from 3 heterogeneous sources to optimize bi-encoder training:

\begin{enumerate}
    \item \textbf{Graded relevance label:} a 5-point relevance rating $r(q,d)\in\{0,1,2,3,4\}$ produced by a cascade of cross encoder teacher models and available human annotations.
    \item \textbf{Multi-channel retrieval prior score} derived from the rank and consensus of retrieved items across multiple existing  production retrieval channels
    \item \textbf{Historical user engagement:} aggregated debiased interaction outcomes for every QIP used as a preference signal to refine supervision for semantically relevant QIPs.
\end{enumerate}

These sources capture complementary dimensions of retrieval quality: (i) graded relevance signals provide the fine-grained semantic supervision for nuanced understanding; (ii) retrieval prior score encapsulates the production system's failure modes and success patterns. This facilitates hard-negative mining by identifying highly-ranked but
irrelevant candidates while reinforcing highly ranked and relevant positive items; (iii) user engagement aligns results with user preferences across semantically relevant items. Highly engaged items implicitly encapsulate tangential item features(other than relevance) including items with active promotions, low prices, faster shipping, strong seller and review ratings, and items which honor user search query intent (brand, color, size, and dietary preference match to query)

\subsection{Graded relevance label}
\label{sec:msd_ce}
We annotate each QIP with an ordinal relevance rating $Rel(q,i)\in\{0,1,2,3,4\}$, corresponding to
$0$ (Embarrassing), $1$ (Bad), $2$ (Okay), $3$ (Good), and $4$ (Excellent). This scale represents the degree of intent fulfillment, ranging from fundamental category mismatches (\(0\)) to precise, ideal matches (\(4\)) that satisfy all query constraints.
Ratings are produced by a combination of available human annotations and a cascade of relevance models (fine-tuned on internal data) evaluated sequentially from least to most computationally expensive
(Gemma-1B, Gemma-2B, and a LLaMA-3 8B model). Each stage outputs a 5-class predictive distribution, and a prediction is accepted early when the confidence score exceeds a stage-specific threshold. For pairs not accepted early, we run all stages, take the majority label, and break ties using the final-stage prediction. We map the relevance rating to a normalized score in $[0,1]$. 
\begin{equation}
\small
\label{rel_score_eqn}
rel\_score(q,i)=(Rel(q,i)-2)/2
\end{equation}


\subsection{Multi channel retrieval prior score}
\label{sec:msd_rank}
We obtain candidates from multiple retrieval channels. Let $\mathcal{S}$ denote the set of retrieval channels, with $|\mathcal{S}|=3$ in our setting.
For each channel $s\in\mathcal{S}$, we record the rank position $r_s(q,i)$ of item $i$ for query $q$ whenever $i$ is retrieved by $s$. Since different channels have different rank ranges and retrieval characteristics, we use a
source-specific normalization constant ${R}_s$ (a maximum rank considered for that channel) and map ranks to a bounded, monotonic prior score:
\begin{equation}
\small
\pi_s(q,i) \;=\; \max\!\left(0,\; 1 - \frac{\log(\max(1,r_s(q,i)))}{\log({R}_s)}\right)\in[0,1].
\end{equation}
When an item is retrieved by multiple channels, we aggregate priors across channels: 
$\pi(q,i) \;=\; \max_{s:\,i\in \mathcal{R}_s(q)} \pi_s(q,i)$,
where ${\mathcal{R}}_s(q)$ denotes the ranked list returned by channel $s$ for query $q$. Higher-ranked items receive larger prior scores: for negatives, this prior serves as a difficulty signal for hard-negative mining, while for positives it reinforces high-confidence behavior already exhibited by the production system. We also compute a multi-channel agreement score
$\mathcal{C}(q,i) \;=\; \frac{c(q,i)}{|\mathcal{S}|}$,
where $c(q,i)$ is the number of channels that retrieved $i$ and $|\mathcal{S}|$ is the total number of channels. We detail how $\mathcal{C}(q,i)$ and $\pi(q,i)$ are combined with the other signals when constructing the unified QIP scores in Section~\ref{sec:qip_scoring}.

\subsection{User engagement signals}
\label{sec:msd_eng}
For each QIP $(q,i)$ we compute an engagement score from aggregated behavioral counts: orders $O(q,i)$, add-to-cart events $A(q,i)$,
clicks $C(q,i)$, and views $V(q,i)$. We form a weighted, log-compressed signal
\begin{equation}
\small
E_{\text{raw}}(q,i) \;=\; \log\!\Big(1 + \lambda_{1}\,O(q,i) + \lambda_{2}\,A(q,i) + \lambda_{3}\,C(q,i) + \lambda_{4}\,V(q,i)\Big),
\end{equation}
where $\lambda_{1}, \lambda_{2}, \lambda_{3}, \lambda_{4}$ are finetuned on validation set (1.5, 0.3, 0.1, and 0.01 were optimal values found through cross-validation). We normalize within each query by the maximum over its candidate set $\mathcal{C}(q)$,
\begin{equation}
\small
E(q,i) \;=\; \frac{E_{\text{raw}}(q,i)}{\max_{d'\in\mathcal{C}(q)}E_{\text{raw}}(q,i') + \epsilon}\in[0,1],
\end{equation}
and apply a sigmoid smoothing (centered at $0.5$)
\begin{equation}
\small
\tilde{E}(q,i) \;=\; \sigma\!\Big(k\,(E(q,i)-0.5)\Big), \qquad \sigma(x)=\frac{1}{1+e^{-x}}, \quad k=8.
\end{equation}
We incorporate engagement primarily for
semantically relevant pairs to avoid promoting popular but completely irrelevant items.
\subsection{Unified supervision and QIP scoring}
\label{sec:qip_scoring}
We assign score to relevant and irrelevant QIPs differently. First, we leverage the graded relevance rating to label the QIPs as either relevant or irrelevant: QIPs with $Rel(q,i)\in\{3,4\}$ are treated as positives, and QIPs with $Rel(q,i)\in\{0,1,2\}$ are treated as negatives. We then assign continuous scores to guide training and sampling. 
\begin{table}[t]
\centering
\small
\setlength{\tabcolsep}{3pt}
\renewcommand{\arraystretch}{1.1}
\resizebox{\columnwidth}{!}{%
\begin{tabular}{l c c c}
\hline
\textbf{Metric} & \textbf{Control (prod)} & \textbf{Var-1 (Rel-only)} & \textbf{Var-2 (Rel+Eng)} \\
\hline
Avg relevance@25 & 3.040 & 3.263 (\,+7.3\%\,) & 3.277 (\,+7.8\%\,) \\
P@25 & 0.794 & 0.873 (\,+10.0\%\,) & 0.877 (\,+10.5\%\,) \\
NDCG@25 & 0.867 & 0.913 (\,+5.4\%\,) & 0.916 (\,+5.7\%\,) \\
\hline
\end{tabular}%
}
\caption{Offline retrieval evaluation at $K{=}25$. Relative gains are computed against Control (prod).The top-25 list contains more relevant items on average for both the variations.}
\label{tab:offline_quality_k25}
\vspace{-1.8em}
\end{table}
\textbf{Positive QIP relevance score}:
We combine the normalized relevance score defined in Eqn. \ref{rel_score_eqn} with the aggregated rank-prior score
$\pi(q,i)$ and multi-channel consensus  score $\mathcal{C}(q,i)$ defined in section \ref{sec:msd_rank}.
\begin{figure*}
    \centering
    \includegraphics[width=0.90\linewidth]{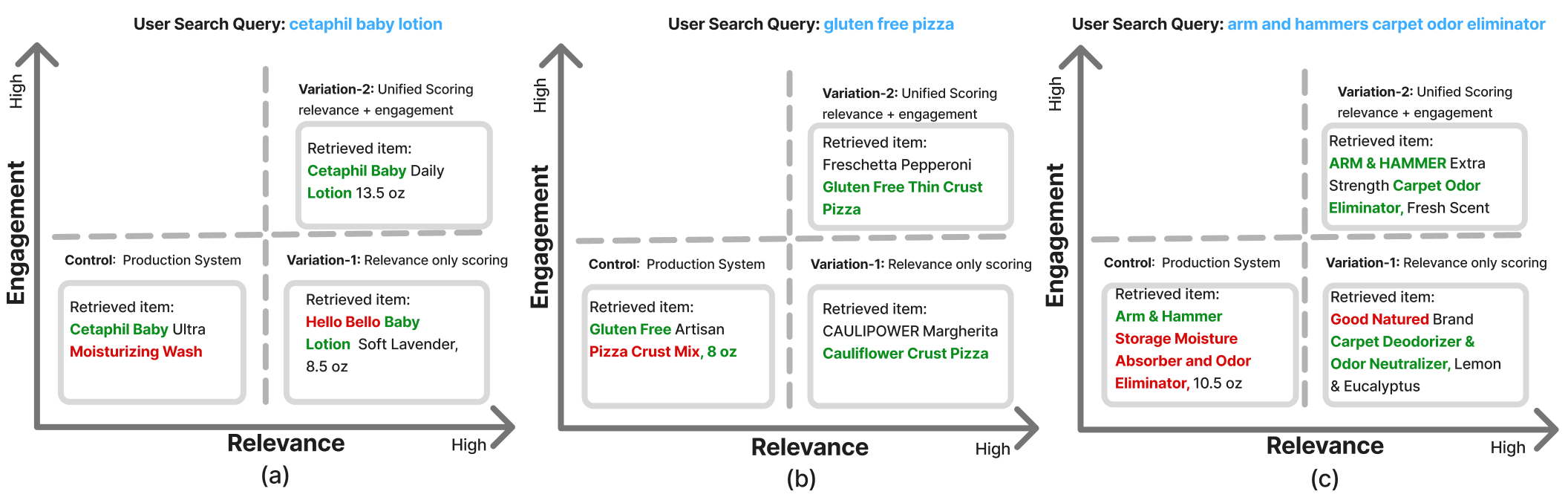}
    \caption{Qualitative results: Green text highlights phrases that align with user intent, while red text shows failure cases. Unified supervision (Variation-2) retrieves relevant and popular(top 1\% engaged) items for each query compared to production system (Control), while relevance only scoring (Variation-1) retrieves relevant but unpopular(bottom 10\% engaged) items.
     }
    \label{fig:qualitative_ttbv3_sigir}
\end{figure*}
We define the unified relevance-rank score as:
\begin{equation}
\small
\label{rel_only_scoring}
y_{\text{rel-rank}}(q,i) = \alpha \cdot rel\_score(q,i) + \beta \cdot \pi(q,i)  + \gamma \cdot \mathcal{C}(q,i) 
\end{equation}

where $\alpha, \beta, \gamma$ (0.6, 0.3, and 0.1  were optimal values found through cross validation) are weights controlling the contribution of the semantic label, rank-prior, and channel consensus. This enables the bi-encoder to reinforce established system strengths through retrieval prior and channel agreement score while learning semantic nuances through relevance score. We then define the engagement-augmented supervision target by adding an engagement boost $\tilde{E}(q,i)$ (defined in Section~\ref{sec:msd_eng}) to the fused relevance--rank score:
\begin{equation}
\small
\label{rel_unified_score}
y_{\text{rel-rank-eng}}(q,i) = \mathrm{clip}\left(\mu_{\text{rel}} \cdot y_{\text{rel-rank}}(q,i) + \lambda_{\text{eng}} \cdot \tilde{E}(q,i), 0, 1\right)
\end{equation}

where $\mu_{\text{rel}}\ge 0$ and $\lambda_{\text{eng}}\ge 0$ controls the strength of the relevance and engagement boost respectively and is tuned on validation(optimal values obtained were 0.85 and 0.15 respectively).
\begin{figure}
    \centering
    \includegraphics[width=\columnwidth]{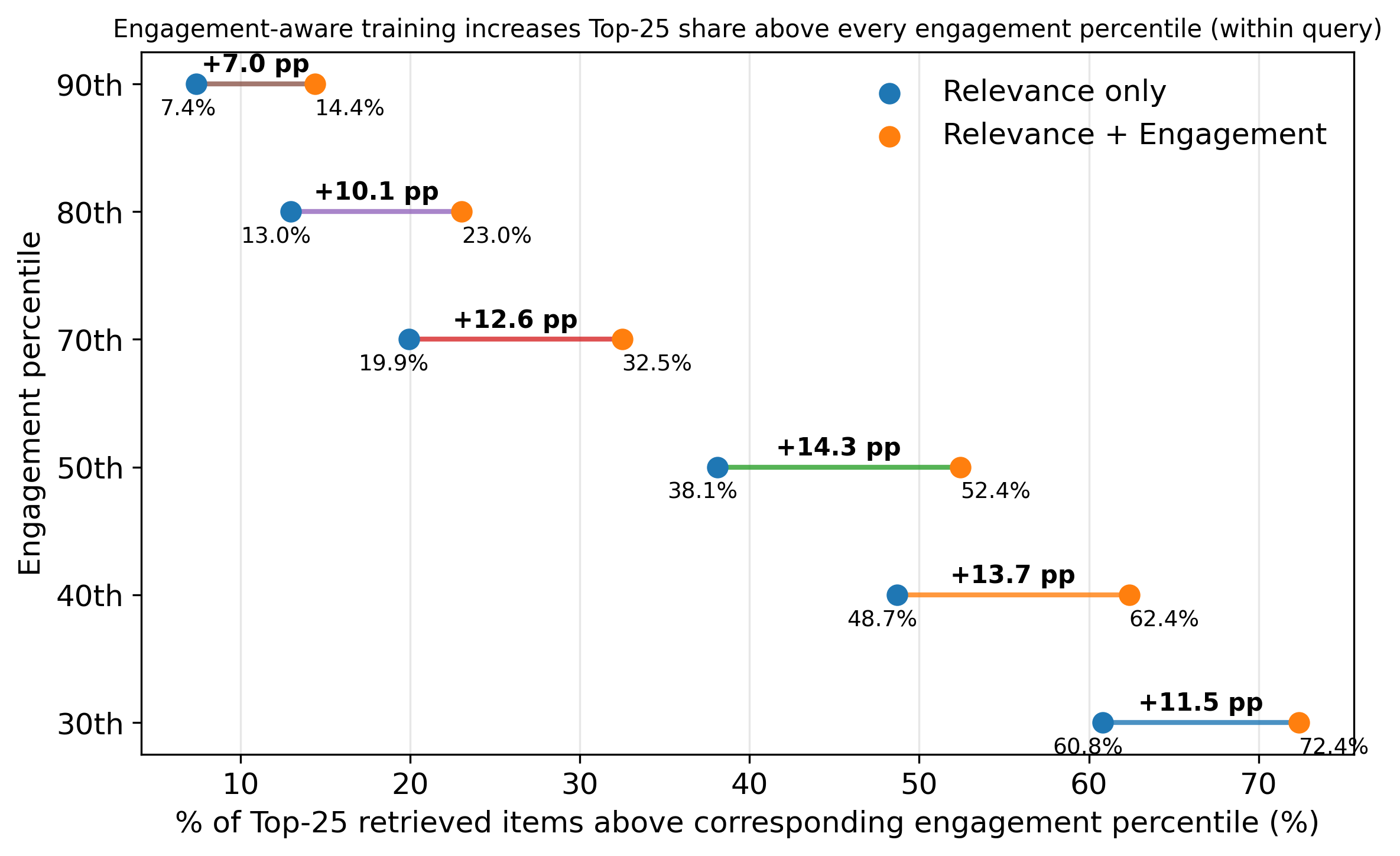}
    \caption{Engagement supervision increases the share of highly engaged items in the Top 25 retrieved set without degrading relevance. For each query, we convert engagement to a within-query percentile and report the percentage of Top-25 items above each percentile cutoff (y-axis). Labels report absolute gains in percentage points (pp)}
    \label{fig:engagement_density_top25}
\end{figure}

\textbf{Irrelevant QIP Difficulty Score}:
For irrelevant pairs ($R \le 2$), we define a difficulty score that prioritizes negatives based on two criteria: retrieval prior and lexical overlap. retrieval prior score $\pi(q,i)$ to identify highly-ranked false positives. We use token similarity to identify negatives that have high keyword overlap with the query. 
\label{irrelevant}
\begin{equation}
\small    
\label{irrelevant_score}
y_{\text{irrel-rank-lex}}(q,i) = \kappa_1 \cdot \pi(q,i) + \kappa_2 \cdot \text{token\_sim}(q,i)
\end{equation}
This facilitates hard-negative mining by forcing the bi-encoder to focus on the most challenging negatives and address existing system's failure modes.

\subsection{Retrieval Model Training}
\label{sec:model_training}

A MiniLM bi-encoder\cite{reimers-2019-sentence-bert} is fine-tuned using the unified score defined in Equation \ref{rel_unified_score} via Cosine Similarity Loss and contrastive learning via Cached Multiple Negatives Ranking (cachedMNR) Loss \cite{henderson2017efficient, gao2021scaling}. 
We use the irrelevant QIP difficulty scores (eq. \ref{irrelevant_score}) in curriculum-based weighting sampling and then use normalized relevance score (Eqn. \ref{rel_score_eqn}) as negative QIP score target for cosine loss.


%% file: Sections/Evaluation.tex
\vspace{-0.4em}
\section{Evaluation}
\label{sec:eval}

\subsection{Evaluation Protocol}
\label{sec:eval_protocol}

\paragraph{Offline evaluation}
We construct an evaluation set of queries that covers both head and tail traffic regimes. Queries are sampled from (i) high-traffic segment, (ii) high-revenue segments, (iii) long-tail queries, and (iv) historically under-performing queries. We build an item index over the full item inventory using FAISS \cite{johnson2019billion} and retrieve the top-$25$ candidates for each query. We report metrics at $K{=}25$ because the sponsored search page has a fixed number of ad slots; in our setting, $25$ is the optimal number of ad items for our application. Each retrieved query--item pair is annotated on a 5-point relevance scale. We use a hybrid judging setup consisting of (i) available human annotations and (ii) model-based annotations from a strong DeBERTa cross-encoder relevance model fine tuned on internal data for pairs without human labels. 
We use the resulting 5-point labels to compute graded retrieval metrics. We report (i) average relevance score over the retrieved list, (ii) Precision@$K$, and (iii) NDCG@$K$
\begin{table}[t!]
\centering
\small
\begin{tabular}{lcc}
\hline
\textbf{Metric} & \textbf{Lift} & \textbf{$p$-value} \\
\hline
\textbf{Sponsored Ad Impressions} & \textbf{+0.60\%} & \textbf{0.03} \\
\textbf{Sponsored Ad Views} & \textbf{+0.49\%} & \textbf{0.09} \\
Sponsored Ad Revenue & +0.45\% & 0.33 \\
\textbf{Add to Cart Rate} & \textbf{+0.99\%} & \textbf{0.009} \\
GMV & +0.37\% & 0.63 \\
Conversion Rate & +0.13\% & 0.64 \\
\textbf{Total Search Page Views per Session} & \textbf{+0.63\%} & \textbf{0.03} \\
\textbf{Total Cart Page Views per Session} & \textbf{+0.87\%} & \textbf{0.02} \\
\hline
\end{tabular}
\caption{A/B Test Results: Business and Ad Engagement Metrics (statistically significant metrics highlighted).}
\label{tab:ab_results}
\vspace{-1.0em}
\end{table}
\emph{Online evaluation.}
Table~\ref{tab:ab_results} summarizes the A/B test results across ad engagement and business metrics. We observe statistically significant results across several metrics with overall positive trend across all metrics.


\subsection{Quantitative Results}
\label{sec:results}

We evaluate all models on the same test set of 30{,}303 queries. Relevance is measured using the 5-point graded ratings from Section~\ref{sec:msd_ce}. Table~\ref{tab:offline_quality_k25} reports quality at $K{=}25$.

\textbf{Relevance-only scoring vs. Control:} 
Relevance-only scoring uses only grade relevance and rank priors (Equation \ref{rel_only_scoring}). Precision rises from 0.794 to 0.873 (+10.0\%) and NDCG rises from 0.867 to 0.913 (+5.4\%). Avg relevance@25 increases from 3.040 to 3.263 (+7.3\%).

\textbf{Adding engagement signal} (Equation \ref{rel_unified_score}) further improves the top-25 list without changing the relevance definition. Relative to the production control, Rel+Eng achieves 0.877 P@25 (+10.5\%), 0.916 NDCG@25 (+5.7\%), and 3.277 Avg relevance@25 (+7.8\%). The gains over Rel-only are modest but consistent, suggesting engagement helps prioritize stronger candidates \emph{among} already relevant items rather than promoting popular results. 

Figure~\ref{fig:engagement_density_top25} measures the percentage of highly-engaged items is present in the retriever's Top-25 list.
We first convert engagement into a within-query percentile (so the 90th percentile means “top 10\% most-engaged items for that query”), then measure what fraction of the Top-25 exceeds each cutoff. The x-axis reports that fraction as a percentage, while the y-axis represents the engagement percentile threshold. Across all thresholds, the engagement supervised model retrieves a higher share of high-engagement items than the relevance-only supervised model.
For example, at the 50th percentile threshold, the share increases by 14.3 pp, and at the 90th percentile threshold it increases by 7.0 pp. 

\vspace{-0.4em}
\subsection{Qualitative Results}
\label{sec:qual}
In Figure \ref{fig:qualitative_ttbv3_sigir}, both variations retrieve relevant items over control. However, the items retrieved by the unified supervision(variation-2) approach are popular too. This item popularity(measured by high user engagement) stems from those items having promotional offers, faster shipping, SNAP-EBT eligible, competetively priced, sold by highly-rated sellers, strong reviews or have high match the user query intent(size, dietary preference, brand, etc.)

%% file: Sections/Discussion.tex
\section{Discussion}
\label{sec:conclusion}

We presented a unified supervision framework for bi-encoder retrieval in Walmart sponsored search that makes semantic relevance the primary training signal and uses engagement only to refine preferences among relevant items. We also incorporate a retrieval-prior signal from production channels to surface hard negatives (highly ranked but irrelevant candidates) and reinforce strong positives. Our results support a simple principle: use relevance to define what is eligible to retrieve, then use engagement to refine ordering among relevant candidates to better align retrieval with downstream ranking objectives.

